\documentclass[a4paper]{jpconf}
\usepackage{graphicx}
\usepackage{amsmath}
\usepackage{amsfonts}
\usepackage{float}
\def\d{\mathrm{d}\hspace*{-0.1ex}}

\begin{document}
\title{Simulating Met-Enkephalin With Population Annealing Molecular Dynamics}

\author{Henrik Christiansen$^1$, Martin Weigel$^2$, Wolfhard Janke$^1$}
\address{$^1$Institut f\"ur Theoretische Physik, Universit\"at Leipzig, IPF 231101, 04081 Leipzig, Germany\\
  $^2$Centre for Fluid and Complex Systems, Coventry University, Coventry, CV1 5FB, England
}

\ead{\\
  henrik.christiansen@itp.uni-leipzig.de\\
  martin.weigel@coventry.ac.uk\\
  wolfhard.janke@itp.uni-leipzig.de\\
}

\begin{abstract}
  Met-enkephalin, one of the smallest opiate peptides and an important
  neuro\-transmitter, is a widely used benchmarking problem in the field of molecular
  simulation. Through its range of possible low-temperature conformations separated
  by free-energy barriers it was previously found to be hard to thermalize using
  straight canonical molecular dynamics simulations. Here, we demonstrate how one can
  use the recently proposed population annealing molecular dynamics scheme to
  overcome these difficulties. We show how the use of multi-histogram reweighting
  allows one to accurately estimate the density of states of the system and hence
  derive estimates such as the potential energy as quasi continuous functions of
  temperature.  We further investigate the free-energy surface as a function of
  end-to-end distance and radius-of-gyration and observe two distinct basins of
  attraction.
\end{abstract}

\section{Introduction}

The simulation of biomolecular systems with molecular dynamics and related techniques
is now a field of vast importance, including applications of paramount fundamental
relevance such as protein folding \cite{scheraga:07} as well as crucial practical
challenges such as those encountered, for instance, in drug discovery
\cite{kitchen:04}. The dominant difficulty in this context results from the slow
relaxation of such systems in the standard (micro-)canonical dynamics, that typically
is a consequence of the presence of several minima in the free-energy landscape
\cite{janke:07}. In the past 35 years or so, an array of tools designed to take on
this arduous task have been developed. In Monte Carlo, the concept of {\em
  generalized ensembles\/} has led to a number of successful approaches, including
simulated tempering \cite{marinari:92a}, parallel tempering
\cite{geyer:91,hukushima:96a}, and multicanonical simulations
\cite{berg:92b}. Parallel tempering or replica exchange uses several copies of the
system run at different temperatures, and conformation exchanges of copies at nearby
temperatures allow for the efficient exploration of the free-energy landscape through
the escape to a regime of fast relaxation. While initially proposed for Monte Carlo,
this successful meta-algorithm was subsequently also adapted to molecular dynamics
\cite{hansmann:97,sugita:99}. For the latter, a number of alternative strategies for
accelerated simulations have been proposed, including so-called ``accelerated''
simulations \cite{hamelberg2004accelerated} as well as metadynamics
\cite{laio:02}. These, and related techniques, are based on the idea of adaptively
biasing the weight function to gradually overcome energy barriers and enable the
sampling of a wide range of the relevant reaction coordinates. In this sense, these
techniques are akin to the multicanonical \cite{berg:92b} and Wang-Landau
\cite{wang:01a} methods in the world of Monte Carlo simulations.

While these methods can significantly extend the degree to which the free-energy
landscape is being explored and accelerate the convergence to equilibrium,
simulations of relevant biopolymers still frequently require very substantial
computational resources. The significant growth in the number of cores in available
(super-)computers is unfortunately only of limited utility to simulations using these
techniques, as exploration of phase space and convergence to equilibrium are
intimately linked to the number of time steps, and there is only limited scope for
speeding up each step through (parallel) task splitting \cite{weigel:18}. A natural
way of using the available parallel resources consists of running many short
simulations independently and combining the resulting statistics to improve the
degree of sampling of the relevant states \cite{gross:17}. While this can work
reasonably well, it does not remove the bottleneck of equilibration as each
simulation conducted in parallel must at least run for the time required to
thermalize. The popular use of Markov state models describing transitions between the
valleys \cite{chodera2007automatic, pande2010everything} can only lead to trustworthy
results if a reliable picture of the relevant states has already been deduced from
direct simulations before.

In the present contribution, instead, we demonstrate how a recently introduced
massively parallel approach to molecular dynamics (MD) simulations, population
annealing molecular dynamics (PAMD), can be employed to utilize practically any
amount of parallel resources available to improve the sampling in systems that are
found hard to thermalize
\cite{christiansen:18,christiansen2019population}. Population annealing, that was
first introduced in the context of Monte Carlo simulations
\cite{iba:01,hukushima:03,machta:10a,barash:16}, uses a large population of identical
system copies (replicas) which are successively cooled down starting from a
thermalized population at high temperatures where equilibration is
straightforward. At each cooling step, the population is resampled, favoring copies
well adapted to the lower temperature. This process is reminiscent of genetic
algorithms \cite{holland1992adaptation} where a population is evolved according to
its fitness. For each new generation mutations occur at a prescribed rate, i.e.,
random changes in the microscopic variables. Subsequently, the new resulting fitness
is calculated.  According to some determined threshold, conformations are replicated
or pruned. Genetic algorithms are hugely successful as optimization algorithms, i.e.,
in finding ground-state configurations \cite{deaven1995molecular,beasley1996genetic}.
However, no thermodynamic information can be extracted. On the other hand, PAMD
provides an equilibrium sample at each temperature considered, and it can hence be
seen as an equilibrium variant of a combination of simulated annealing
\cite{kirkpatrick:83} and a genetic algorithm.

\section{Model and Method}

The system studied here is Met-enkephalin, a penta-peptide with amino-acid sequence
Tyr-Gly-Gly-Phe-Met, that occurs in many organisms. It inhibits neurotransmitter
release upon activation of the appropriate opioid receptor, and so plays a central
role in pain regulation \cite{hughes:75}.  It is said to inhibit tumor growth and
metastasis \cite{kuniyasu2010cd10}. For the simulation, we cap the ends with a methyl
and an acetyl group, respectively, resulting in a total size of $84$ atoms. The
interactions were modeled by the AMBER force-field ff94 \cite{cornell:95}.

\begin{figure}[tb!]
  \def \scaling {0.12}
  \centering
  \begin{tabular}{cccc}
    \includegraphics[scale=\scaling]{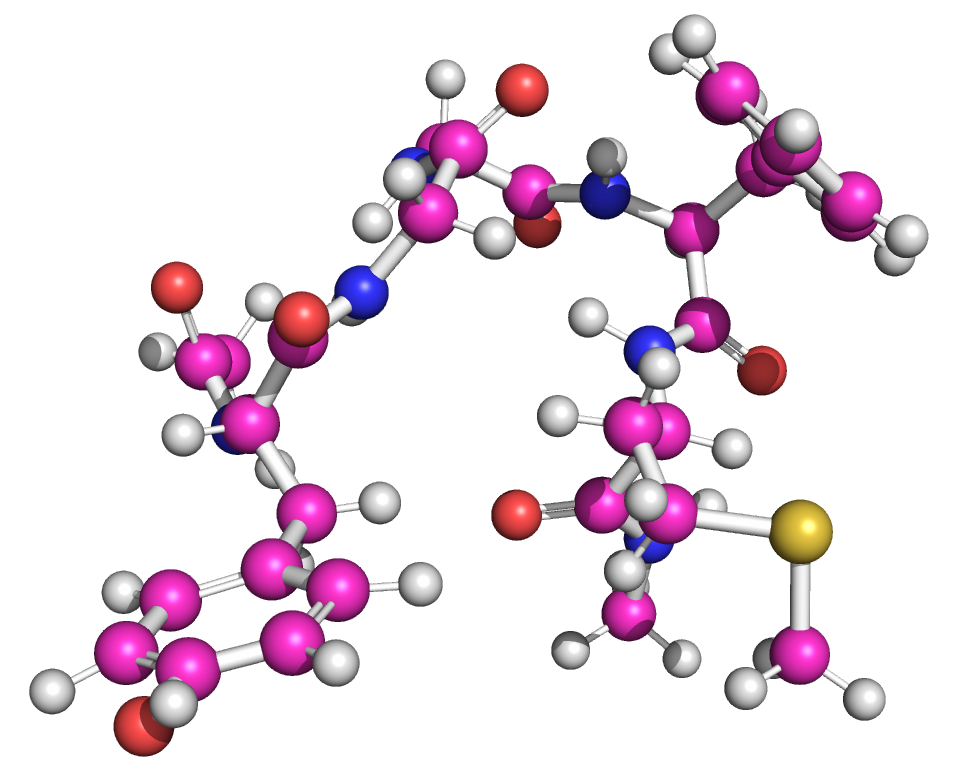}&\includegraphics[scale=\scaling]{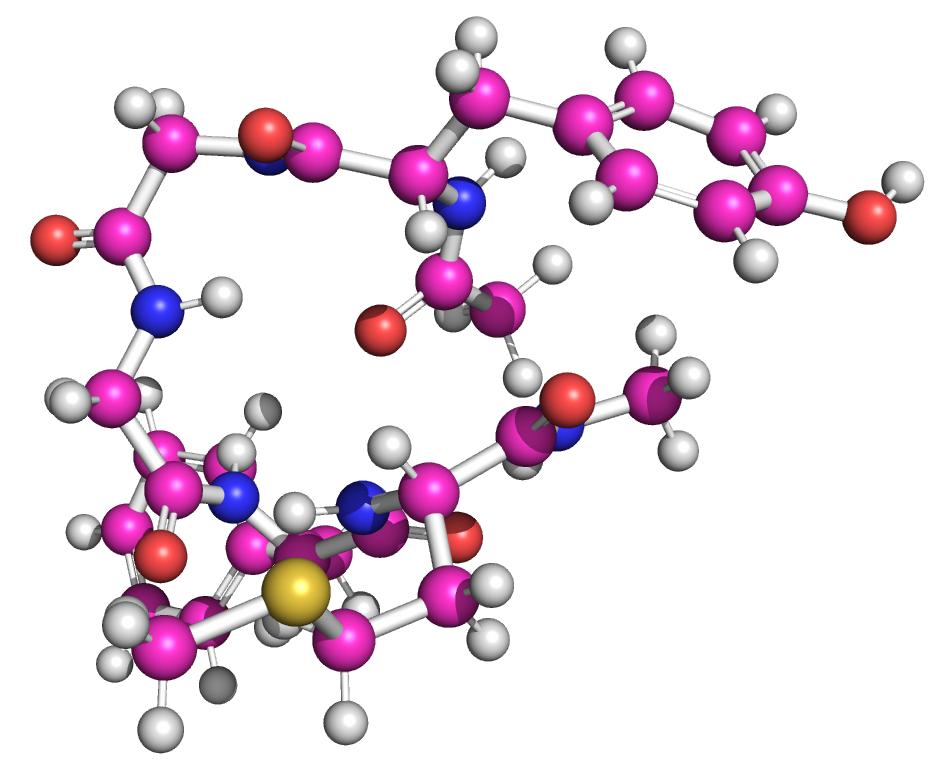}&\includegraphics[scale=\scaling]{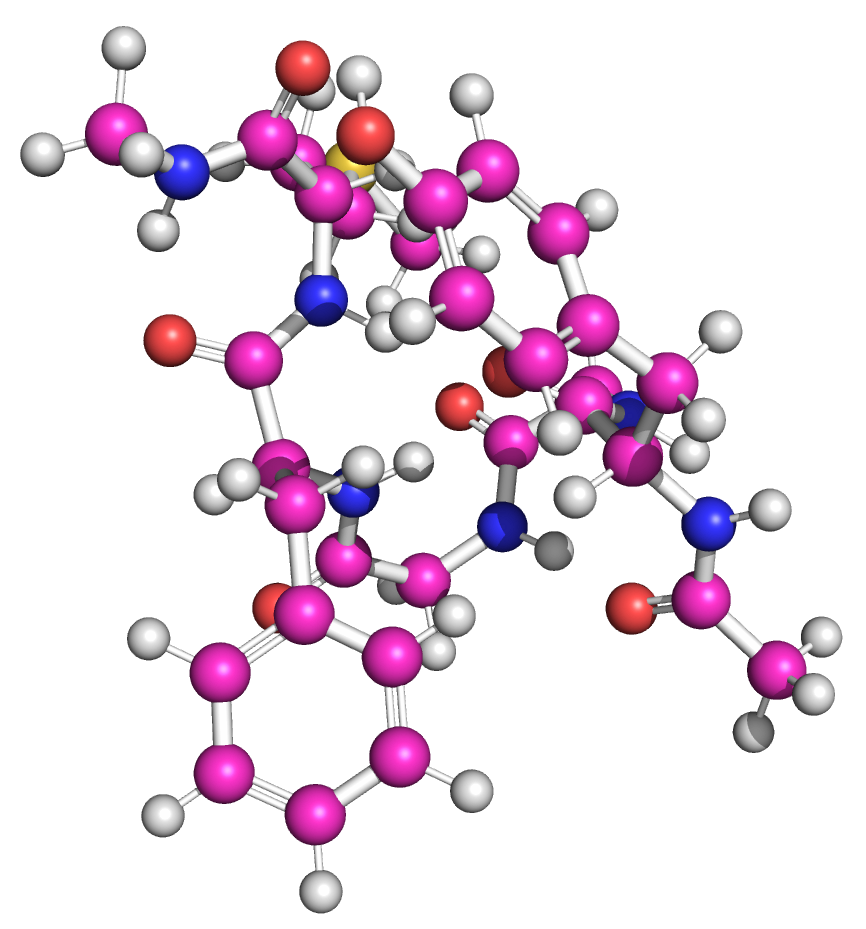}&\includegraphics[scale=\scaling]{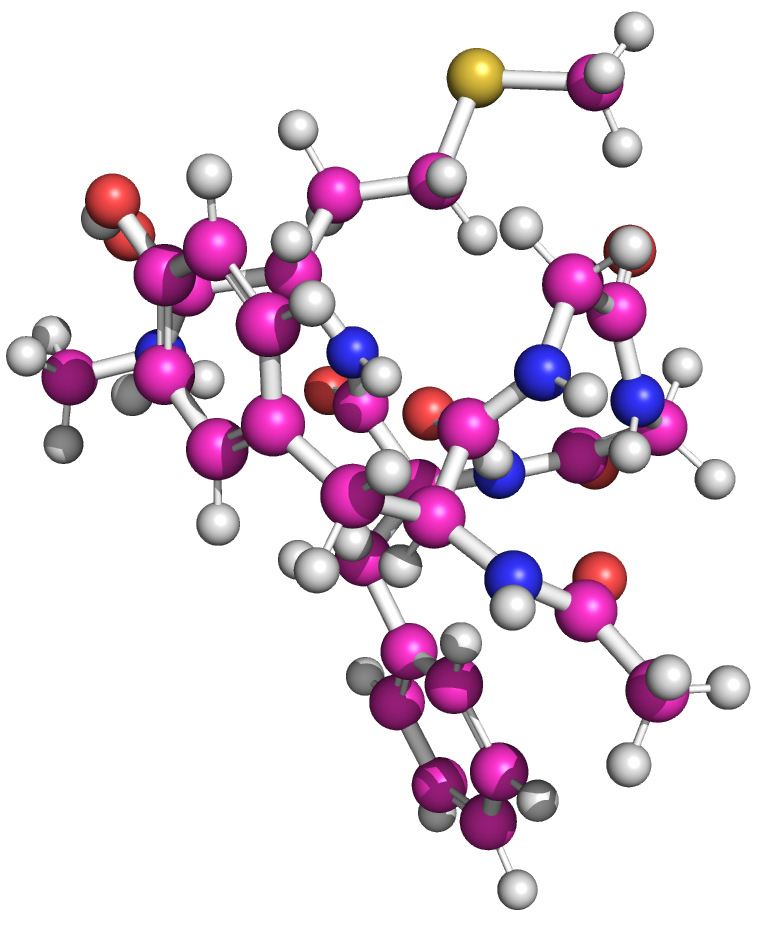}\\
    $T=585$~K&$T=342$~K&$T=286$~K&$T=200$~K\\
    \end{tabular}
    \caption{Snapshots of Met-enkephalin at different temperatures, decreasing from
      left to right. The following color-code applies: pink -- carbon, white -- hydrogen,
      red -- oxygen, blue -- nitrogen, yellow -- sulfur.}
  \label{Snapshots}
\end{figure}

While previous simulation studies of this system mostly used the replica exchange
method for ensuring that equilibration is achieved, we here employ PAMD simulations
to study the system \cite{christiansen:18}. In particular, this involves the
following simulation steps:
\begin{enumerate}
\item Set up an equilibrium ensemble of $R$ independent copies of the peptide at some
  high temperature $T_0$.
\item Choose the next temperature $T_i < T_{i-1}$ from a pre-defined sequence.
\item Resample the population of systems to the new temperature $T_i$ by replicating
  each copy a number of times proportional to the relative Boltzmann weight
  $\tau_j \propto e^{-(1/k_BT_i-1/k_BT_{i-1})E_j}$, where $E_j$ is the potential
  energy of the $j$th replica. The momenta are simply adjusted to the new temperature
  by rescaling $p_k \rightarrow \sqrt{\frac{T_i}{T_{i-1}}}p_k$.
\item Update each replica with $\theta$ simulation steps of the underlying MD algorithm.
\item Calculate observables $\mathcal{O}$ at temperature $T_i$ as population averages.
\item Goto step (ii) until $T_i$ reaches or falls below the target temperature $T_N$.
\end{enumerate}
In parallel tempering, the choice of a suitable temperature protocol is often a
challenging task \cite{katzgraber:06,bittner:08}. For PAMD, on the other hand, we
have shown in Ref.~\cite{christiansen2019population} that an appropriate temperature
set may be found ``on the fly'' by demanding a constant energy histogram overlap
without the need of conducting preliminary simulations. For the present study we
adopt the previously used temperature scheme consisting of simulations at $T_i=700$,
$585$, $489$, $409$, $342$, $286$, $239$, $200$~K. This corresponds to a constant
energy histogram overlap of about $30\%$ \cite{christiansen2019population}. Due to
the resampling that involves making copies of some population members, it is crucial
to employ a stochastic thermostat in PAMD simulations, as otherwise the copies would
eternally follow the same trajectory, thus compromising the quality of the
statistical sampling \cite{christiansen:18}. Here we use a Langevin thermostat with
$0.5$~fs integration step and friction coefficient $\gamma=1/$ps. For our simulation
we ran $10^4$ replica with a total simulation time of $200$~ns.  MD steps amounting
to $25$~ns were used to initialize the population at the highest temperature,
followed by $21.875$~ps of evolution (or $4375$ updates) performed at each
temperature step. For the molecular dynamics part, which dominates the computational
effort, we relied on the package OpenMM \cite{eastman2017openmm}.  Overall, our
simulations show an excellent parallel efficiency of over $85\%$ using $500$ cores,
far superior to what normally can be achieved in parallel tempering.

\section{Results}

To convey an overall impression of the behavior of Met-enkephalin, we present typical
snapshots of the peptide in Fig.~\ref{Snapshots}. It is clearly seen that compared to
the initial configuration at the highest temperature (left), the conformations
obtained by population annealing at lower temperatures (to the right) are
significantly more compact.
%We did not attempt to extract information about the
%ground state of the system from our simulations, but it is clear that this should be
%possible with a suitably chosen final temperature \cite{hansmann:97}.

\begin{figure}[tb]
  \centering
  \includegraphics{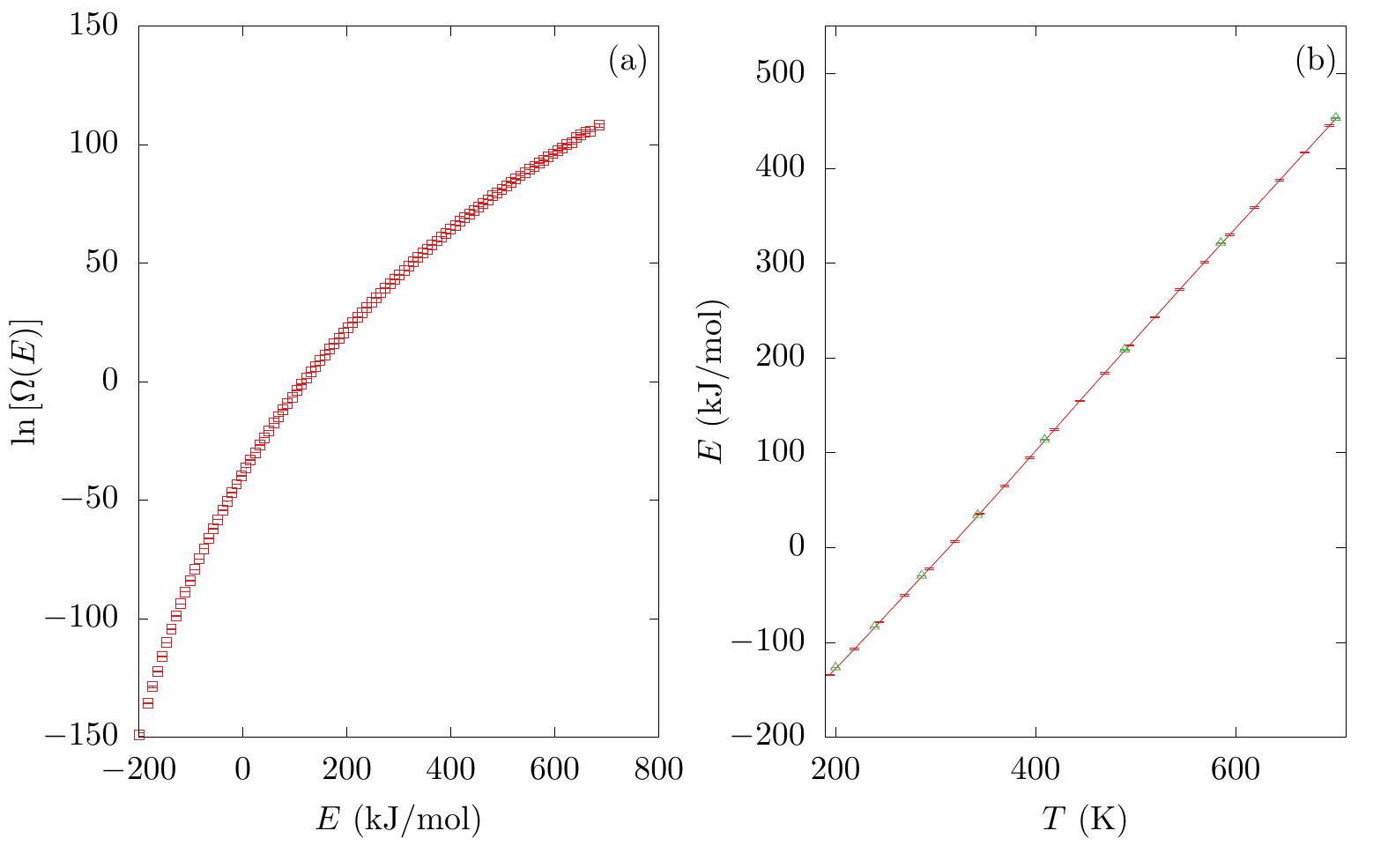}
  \caption{(a) Logarithm of the density of states for Met-enkephalin obtained from
    our PAMD simulation using WHAM \cite{ferrenberg:89a,kumar:92,barash:18}. (b)
    (Potential) energy $E$ as derived from the density of states of (a) as a function
    of temperature $T$. The green triangles indicate population averages at the
    actual simulation temperatures.}
  \label{DOS}
\end{figure}

Thermodynamic properties at the temperatures included in the simulation protocol can
be extracted from regular population averages. Beyond that, effectively continuous
estimates of observables can be derived from the density of states $\Omega(E)$ that
can be obtained using the weighted-histogram analysis method (WHAM)
\cite{ferrenberg:89a,kumar:92} applied to the potential energy $E$. Details of the
application to population annealing can be found in Ref.~\cite{barash:18}. The result
of this analysis for Met-enkephalin is shown in Fig.~\ref{DOS}(a). Note that due to
the continuous nature of the energy functional one needs to use binning to apply
WHAM. Here, we used $200$ equally spaced bins in the range from the lowest energy
observed to the highest. (The binning is only applied after the simulation, in the
post-processing of data.)  Using this result for the density of states, it is then
possible to extract observable estimates at any temperature such as
\begin{equation}
  \langle E \rangle(T) = \frac{\sum_E E \Omega(E) e^{-E/k_BT}}{\sum_E\Omega(E)e^{-E/k_BT}}.
\end{equation}
The resulting plot is shown in Fig.~\ref{DOS}(b) for temperatures in the simulation
range of $200$~K $<T< 700$~K.

The spatial structure of the peptide can be characterized using quantitative measures
such as the end-to-end distance,
\begin{equation}
  R_\mathrm{ee}=| \mathbf{R}_N - \mathbf{R}_1 |,
\end{equation}
where $\mathbf{R}_N$ is the position of the last atom of the peptide and
$ \mathbf{R}_1$ that of the first.  For a peptide it is not particularly useful to
apply this definition directly and, instead, we rather use the distance between the
two outermost carbon atoms.  While it is known that this quantity can be misleading,
e.g., when ``by chance'' the start and end monomer in an otherwise extended
conformation are very close to each other, it may provide useful insight when
directly compared to other physical quantities.  A more robust indicator of the
average size is the (squared) radius-of-gyration,
\begin{equation}
  %% R_\mathrm{g}^2= N^{-1} \sum_{i=1}^{N} (\mathbf{R}_i-\mathbf{R}_{\mathrm{com}})^2,
  R_\mathrm{g}^2=  \frac{\sum_{i=1}^{N} m_i(\mathbf{R}_i-\mathbf{R}_{\mathrm{com}})^2}{\sum_{i=1}^{N} m_i},
\end{equation}
where the sum from $i=1$ to $N$ runs over all atoms and $\mathbf{R}_{\mathrm{com}} = \sum_i m_i\mathbf{R}_i/\sum_i m_i$ is the center-of-mass position of the peptide.
%% ({\tt does it make sense to use the mass weighting here as in
%% the formal radius-of-gyration definition?}{\bf good catch! Since I originally only used the backbone alpha carbons, this made no difference. Now, correctly, one should include the weighting by mass}).
This observable thus averages the squared distance of each atom weighted by its mass to the center-of-mass and is hence a well defined measure for the current extension of the peptide.
%% We here use all atoms to calculate this observable ({\tt why does that need to be stated? what is the alternative?}{\bf Common alternative are all backbone atoms or the alpha carbons of the backbone.}).

\begin{figure}[tb]
  \centering
  \includegraphics{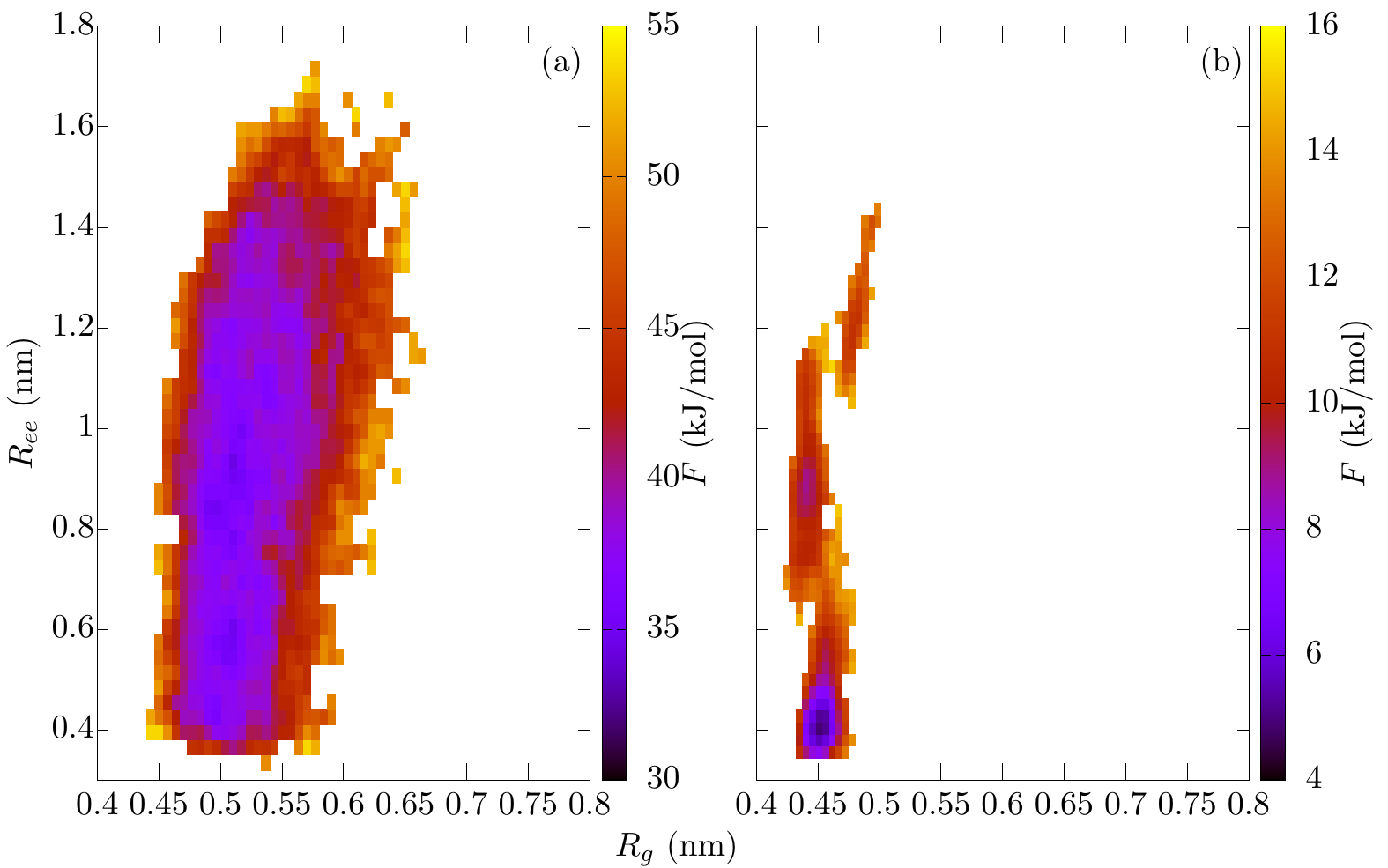}
  \caption{Plots of the end-to-end distance versus the radius-of-gyration for our simulation of Met-enkephalin at (a) $T=700$~K and (b) $T=200$~K.}
  \label{ReRg}
\end{figure}

While it can already be useful to consider averages and potentially higher
moments of quantities such as $R_\mathrm{ee}$ and $ R_\mathrm{g}$, it is even more
instructive to consider their full probability distributions. In general, define
the probability density of an observable $\mathcal{O}$ as
\begin{equation}
  P(\mathcal{O}) \sim \iint \d \mathbf{q} \d \mathbf{p}\,P(\mathbf{q},\mathbf{p})
  \delta\left[\mathcal{O}(\mathbf{q},\mathbf{p})-\mathcal{O}\right],
\end{equation}
where $\mathbf{q}$ are the positions and $\mathbf{p}$ the momenta of the atoms.  It
is then sometimes useful to analyze the {\em free-energy surface\/} 
\begin{equation}
  F(\mathcal{O})= -k_B T \ln \left[ P(\mathcal{O}) \right] + c.
\end{equation}
The constant $c$ is arbitrary and chosen to be zero here.  This signifies the
observation that only free-energy differences are significant, as the normalization
is unknown from simulations.

As an illustration, we plot in Fig.~\ref{ReRg} the free-energy surfaces of the
end-to-end distance versus the radius-of-gyration for (a) the
highest temperature of $700$~K and (b) the lowest temperature of $200$~K.  As
expected, both the end-to-end distance and radius-of-gyration are significantly
smaller for lower temperatures, where the peptide is in a more compact conformation.
At $T=200$~K the free-energy landscape splits in two minima separated by a region in which no
conformations are observed.
%% Also apparent is the splitting in two minima separated by a region in which no
%% conformations are observed for $T=200$~K ({\tt is that apparent?}{\bf maybe not too clear...}).
This signifies two different folding states, which using canonical simulations would not have been
observable in a single run.

\section{Conclusion}

We have described population annealing for molecular dynamics and performed
simulations for the penta-peptide Met-enkephalin, illustrating the utility of this
approach for efficiently thermalizing the system and estimating thermodynamic
quantities. With the help of the weighted-histogram analysis method it is possible to
get reliable estimates of the density of states that in turn enable the derivation of
quasi continuous estimates of observables as a function of temperature. This peptide
has several free-energy minima, which we present by plotting the free-energy surface
of the end-to-end distance versus the radius-of-gyration. Here, at least two basins
of attractions are observed, indicating the presence of a free-energy barrier at
$T=200$~K. In the population simulation method employed here, at low temperatures
different basins are occupied by replicas according to their statistical weight, thus
ensuring a fair sampling of the full free-energy landscape while allowing the
efficient use of massively parallel computational resources.

\ack This project was funded by the Deutsche Forschungsgemeinschaft (DFG, German
Research Foundation) under Grant No.\ 189\,853\,844 -- SFB/TRR 102 (project B04), and
further supported by the Deutsch-Franz\"osische Hochschule (DFH-UFA) through the
Doctoral College ``$\mathbb{L}^4$'' under Grant No.\ CDFA-02-07, the EU Marie Curie
IRSES network DIONICOS under Grant No.\ PIRSES-GA-2013-612707, and the Leipzig
Graduate School of Natural Sciences ``BuildMoNa''.

\section*{References}
\bibliography{paper}

\end{document}